\documentclass[11pt]{article}

\usepackage[paper=a4paper,dvips,top=2.0cm,left=2.2cm,right=2.2cm,bottom=2.9cm,footskip=1.4cm]{geometry}
\usepackage{amsfonts,amsmath,amsthm,amssymb}
\usepackage{mathrsfs}
\numberwithin{equation}{section}  
\usepackage[svgnames]{xcolor}
\usepackage{fix-cm}
\usepackage{sectsty}
\usepackage{fancyhdr}
\usepackage{lastpage}
\usepackage{graphicx}
\usepackage{url}
\usepackage{enumerate}
\usepackage{paralist}
\usepackage{multicol}
\usepackage{color}  
\usepackage{float}  
\usepackage{epsfig}
\usepackage{enumitem}
\usepackage{hyperref}   
\hypersetup{colorlinks=true,linkcolor=beamer@PRD, citecolor=beamer@PRD}
\usepackage{authblk}  
\usepackage{cite}  

\newcommand\myref[1]{\textcolor{beamer@PRD}{(}\ref{#1}\textcolor{beamer@PRD}{)}}

\definecolor{beamer@blue}{RGB}{0,0,255}
\definecolor{beamer@mediumblue}{RGB}{0,0,190}
\definecolor{beamer@midnightblue}{RGB}{25,25,112}
\definecolor{beamer@navy}{RGB}{0,0,128}
\definecolor{beamer@darkblue}{RGB}{0,0,139}
\definecolor{beamer@purple}{RGB}{128,0,128}
\definecolor{beamer@levander}{RGB}{100.,149.,237.}
\definecolor{beamer@PRD}{RGB}{46,48,146}
\definecolor{beamer@green}{RGB}{0,128,0}
\definecolor{beamer@darkgreen}{RGB}{0,100,0}
\definecolor{beamer@olive}{RGB}{128,128,0}
\definecolor{beamer@darkolivegreen}{RGB}{85,107,47}
\definecolor{beamer@gray}{RGB}{190,190,190}
\definecolor{beamer@ivry}{RGB}{220,220,220}
\definecolor{beamer@new}{RGB}{40,120,50}
\definecolor{shadecolor}{RGB}{220,220,220}
\definecolor{beamer@darkslategray}{RGB}{47,79,79}
\definecolor{beamer@chocolate}{RGB}{210,105,30}
\definecolor{beamer@brown}{RGB}{165,42,42}
\definecolor{beamer@orangered}{RGB}{255,69,0}
\definecolor{beamer@maroon}{RGB}{128,0,0}
\definecolor{beamer@white}{RGB}{234,242,243}
\definecolor{beamer@silver}{RGB}{0.5,0.45,0.37}

\begin{document}


\title{\textbf{BTZ black holes in massive gravity}}
\author{\textbf{Sumeet Chougule$^\bullet$, Sanjib Dey$^\ast$, Behnam Pourhassan$^\circ$ and Mir Faizal$^{\dagger\S}$} \\ \small{$^\bullet$School of Physics, University of Hyderabad, Hyderabad, India \\ $^\ast$Department of Physics, Indian Institute of Science Education and Research Mohali, \\ Sector 81, SAS Nagar, Manauli 140306, India \\ $^\circ$School of Physics, Damghan University, Damghan, 3671641167, Iran \\ $^\dagger$Irving K. Barber School of Arts and Sciences, University of British Columbia -Okanagan,\\ Kelowna, British Columbia V1V 1V7, Canada \\ $^{\S}$Department of Physics and Astronomy, University of Lethbridge, Lethbridge, Alberta, T1K 3M4, Canada \\ E-mail: sumeetchougule@live.com, dey@iisermohali.ac.in, b.pourhassan@umz.ac.ir, mirfaizalmir@googlemail.com}}
\date{}
\maketitle
    	
\thispagestyle{fancy}
\begin{abstract}
We analyze certain aspects of BTZ black holes in massive theory of gravity. The black hole solution is obtained by using the Vainshtein and dRGT mechanism, which is asymptotically AdS with an electric charge. We study the Hawking radiation using the tunneling formalism as well as analyze the black hole chemistry for such system. Subsequently, we use the thermodynamic pressure-volume diagram to explore the efficiency of the Carnot heat engine for this system. Some of the important features arising from our solution include the non-existence of quantum effects, critical Van der Walls behaviour, thermal fluctuations and instabilities. Moreover, our solution violates the Reverse Isoperimetric Inequality and, thus, the black hole is super-entropic, perhaps which turns out to be the most interesting characteristics of the BTZ black hole in massive gravity.   
\end{abstract}	 
\section{Introduction} \label{sec1}
\addtolength{\footskip}{-0.1cm} 
\addtolength{\voffset}{1.2cm} 
Astronomical observations suggest that our universe is expanding, in fact, the observational data from the type-I supernovae proposes that our universe is expanding in an accelerating rate \cite{super,super1,super5}. Theoretically, this accelerated expansion is due to the creation of a negative pressure implying a positive vacuum energy density which is caused due to a cosmological constant term in the Einstein's field equation. Such a cosmological constant can be associated with the zero point energy of quantum field theories, however, the value of the cosmological constant obtained from the quantum field theoretical calculations is several orders of magnitude greater than that arising from the observational astronomy. Various models have been proposed to explain the origin of the cosmological constant  \cite{DE,DE1,DE2} including some large distance modifications of general relativity \cite{M1}. All of these modifications are mainly constrained in such a way that they are compatible with the theory of general relativity at a scale at which general relativity has been observed \cite{2a,a2}. Alternative schemes also exist, for instance, it is possible to obtain a long distance infrared modification of the general relativity along with massive  gravitons, where the mass of the graviton can be obtained from the observational data \cite{M2}. It may be noted that by adding a small Fierz-Pauli like mass term to the action of general relativity, we do not obtain a stable theory consistent with the zero mass limit \cite{1}. In fact, such type of straightforward modified theories are not physical as they violate the known experimental bounds obtained from solar system tests \cite{2a,a2}. The inclusion of non-linearities in some theories, where the Vainshtein mechanism helps to recover General Relativity at the solar system scales \cite{4,5}, gives rise to the Boulware-Deser ghost \cite{6}, which can be removed by introducing higher-order terms in the massive action as it was done in the case of the dRGT theory of Massive gravity \cite{14}, where a subclass of massive potentials is considered \cite{7,10}. Within this framework, the higher order-term when grouped with the ghost like term becomes a total derivative and, thus, the action is resumed having two free parameters.

Mass terms can be produced by using many other mechanisms, for instance, by breaking the Lorentz symmetry of the system \cite{lore}. Nevertheless, mass terms are very useful and they have been utilized to study various interesting models; such as, in Gauss-Bonnet massive gravity  \cite{gaus}. It has been observed that the massive gravitons can lead to interesting modification of black hole thermodynamics. The modification to the behavior of black hole by the inclusion of graviton mass has also been analyzed in the extended phase space \cite{M2} in order to study the phase transition of black holes \cite{phase}. Besides, the cosmological solutions \cite{cosm12} and the initial value constraint \cite{cosm14}, etc., have been explored in the context of massive gravity.  
	
BTZ black hole is another interesting object which was introduced \cite{Banados:1992wn}. It is possible to construct a BTZ black hole in massive gravity. In fact, an asymptotically AdS charged BTZ black hole has been constructed in a massive theory of gravity and various different aspects of such a solution have been studied  \cite{main}. In this work, we study both the dynamical and thermodynamical aspects of such a solution. In fact, using the  geometrical thermodynamic approaches Weinhold, Ruppeiner and Quevedo metrics have been studied for this system. We analyze the Hawking radiation \cite{Hawking} for such a system using the tunneling formalism \cite{parikh2000hawking,Kim:2011fh,Yale:2010tn,Nozari:2008rc}. We also analyze the black hole chemistry for this BTZ black hole \cite{Frassino:2015oca,Kubiznak:2016qmn}. This is done by relating  each black hole parameter with a chemical equivalent using the first law of thermodynamics. The cosmological constant is considered a thermodynamic parameter related to the pressure of the system \cite{Frassino:2015oca,Kubiznak:2016qmn}. Subsequently, using the  pressure-volume terms, we study the heat engines and their efficiency \cite{Hennigar:2017apu,Johnson:2014yja}.   
\section{Massive gravity}\label{sec2}
\lhead{BTZ black holes in massive gravity}
\chead{}
\rhead{}
\addtolength{\footskip}{0.1cm} 
\addtolength{\voffset}{-1.2cm} 
Let us start by recollecting the notions of three dimensional massive gravity\cite{14}. First, we define the constants for massive gravity $c_i$ and the symmetric polynomials of the eigenvalues $\mathcal{U}_i$ of the $d\times d$ matrix $\mathcal{K}^\mu_\nu=\sqrt{g^{\mu\alpha}f_{\alpha\nu}}$
\begin{eqnarray}\label{Ui}
&&\mathcal{U}_{1}=[\mathcal{K}],\nonumber\\
&&\mathcal{U}_{2}=[\mathcal{K}]^2 -[\mathcal{K}^2],\nonumber\\
&&\mathcal{U}_{3}=[\mathcal{K}]^3 -3[\mathcal{K}][\mathcal{K}^2]  + 2[\mathcal{K}^3],\nonumber\\
&&\mathcal{U}_{4}=[\mathcal{K}]^4 -6[\mathcal{K}^2][\mathcal{K}]^2  + 8[\mathcal{K}^3][\mathcal{K}]+3[\mathcal{K}^2]^2
-6[\mathcal{K}^4],
\end{eqnarray}
with $[\mathcal{K}]=\mathcal{K}^\mu_\mu$ and $(\sqrt{A})^\mu_\nu(\sqrt{A})^\nu_\lambda=A^\mu_\lambda$. The three dimensional action of massive gravity with an abelian $U(1)$ gauge field and negative cosmological constant is known to have the form \cite{main}
\begin{equation}\label{action}
S=-\frac{1}{16\pi}\int d^{3}x \sqrt{-g}\left[\mathcal{R}-2\Lambda +L(\mathcal{F})+\tilde{M}^{2}\sum_{i=1}^{4} c_i{\cal U}_i(g, f)\right],
\end{equation}
where $L(\mathcal{F})$ is the Lagrangian for the vector gauge field, $\Lambda$ stands for the cosmological constant, $\mathcal{R}$ represents the scalar curvature and $\tilde{M},f$ are the mass term and fixed symmetric tensor, respectively. Also, $\mathcal{F}=F_{\mu\nu}F^{\mu\nu}$ is the Maxwell invariant, with $F_{\mu\nu}=\partial_\mu A_\nu-\partial_\nu A_\mu$ being the Faraday tensor and $A_\mu$ being the gauge potential. By using the action \myref{action}, one can utilize the variational principle to obtain the field equations for gravitation, as computed in \cite{main}. Nevertheless, in order to obtain a static solution of the charged AdS black hole in $3D$, we can start with the following ansatz for the metric  \cite{Cai,main}
\begin{equation}\label{metric1}
ds^2= -f(r)dt^2 +\frac{dr^2}{f(r)}+r^2 d\phi^2,
\end{equation}
where $f(r)$ is an arbitrary function of the radial coordinate. An exact solution of the metric \myref{metric1} can be obtained by choosing a reference metric as given by
\begin{equation}\label{matricAnsatz}
f_{\mu\nu}=\text{diag}(0,0,c^2h_{ij}),
\end{equation}
with $c$ being a positive constant. With the given ansatz \myref{matricAnsatz}, $\mathcal{U}_i$'s can easily be computed as $\mathcal{U}_1=c/r, \mathcal{U}_2=\mathcal{U}_3=\mathcal{U}_4=0$, which indicates that the contribution of massive gravity is arising only from the $\mathcal{U}_1$. Furthermore, keeping in mind that we are going to study a linearly charged BTZ black hole, we can choose the Lagrangian of Maxwell field as $L(\mathcal{F})=-\mathcal{F}$. In addition, by considering a gauge potential related to the radial electric field to be of the form $A_\mu=h(r)\delta_\mu^t$, and by following the procedure explained in \cite{main}, one obtains an exact form of the radial function $f(r)$ as given by \cite{main}
\begin{equation}\label{fr}
f(r)=-\Lambda r^2-m-2q^{2}\ln{(\frac{r}{l})}+\tilde{M}^{2}c c_{1} r,
\end{equation}
from which one obtains the exact expression of the metric for the massive gravity in the given scenario. Here $m = 8M$ and $q = 2Q$, with $M,Q$ being the mass and electric charge of the black hole, respectively. Here, $l$ is an arbitrary constant having the dimension of the length, which is arising from the fact that the logarithmic arguments should be dimensionless. In what follows, we shall consider the cosmological constant $\Lambda=-1/l^2$, since $\Lambda$ has a dimension of inverse squared length. However, it should be noted that the metric corresponding to \myref{fr} can also be constructed by using other methods available in the literature. For instance, in \cite{Arraut1,Arraut2,Houndjo,Arraut3}, the authors have explored a procedure by using the St\"uckelberg method, where the St\"uckelberg fields can be considered to be in a unitary gauge, so that the corresponding fiducial metric becomes the Minkowskian. This is the simplest case that one can consider. However, the fiducial metric coming out of such theories may not be unique, as it depends on the choice of the gauge field. If the St\"uckelberg fields are not in unitary gauge then one obtains an associated fiducial metric also but it is not a Minkowskian anymore. This method is quite simple and, surely, it has its own beauty, however, in this paper, we have constructed the metric from a slightly different procedure as explained earlier in this section, where we have used a reference metric for the purpose.
\section{Hawking Radiation as Tunneling}\label{sec3}
Among many approaches of analyzing the Hawking radiation of a black hole, in this manuscript we consider the method of quantum tunneling \cite{parikh2000hawking,Kim:2011fh,Yale:2010tn,Nozari:2008rc}. There are many reasons for this. For instance, other approaches for deriving Hawking radiation deal with the principle of detailed balance, the background geometry is considered fixed and the energy conservation is not enforced during the emission process in general cases. More precisely, in a general cases like massive gravity, the energy conservation is not valid in the usual sense, however, this is because the time-like Killing vector in massive gravity is not defined in the same direction of the ordinary time-coordinate. Actually, the fact is that the black-hole radiation is related to the way how one defines the time (vacuum), therefore, in order to restore the notion of energy conservation for those cases, one can redefine the time, as indicated in \cite{Arraut1}. There are many ways to redefine such time coordinate, such as, the path integral method \cite{Arraut2,Houndjo}, Bogolibov method \cite{Arraut3}, etc. However, in tunneling formalism the energy conservation is utilized to obtain non-thermal corrections to the spectrum of particles and, thus, this method shows the conservation of energy in a more explicit way. Moreover, as the tunneling process takes place at the horizon, the coordinate system is required to be non-singular at the horizon. Thus, Painlev\'e coordinates are useful for such analysis \cite{painleve}. In this formalism it is argued that when a classical stable system becomes quantum mechanically unstable, it is natural to consider tunneling. The Hawking radiation occurs due to the  tunneling of virtual particles. The idea is to consider the vacuum fluctuations near the horizon which creates a pair of particle and anti-particle. When a pair is created just inside the horizon, the positive energy particle tunnels across the horizon and escapes to infinity as a Hawking radiation. While the black hole absorbs the negative energy particle and its mass is decreased. Similarly, for the pair which is created outside the horizon, the anti-particle tunnels inside the black hole before it is annihilated. Thus, in both of the cases the black hole absorbs the negative energy particle by decreasing its mass, while the positive energy particle escapes to infinity to be observed as Hawking radiation. Under this formalism, the probability of tunneling is given as \cite{parikh2000hawking}
\begin{equation}
\Gamma \sim e^{-2 Im \mathfrak{S}},
\end{equation}
where $\mathfrak{S}$ is the action of the trajectory. The barrier for the tunneling is provided by the outgoing particle itself. Black holes lose energy due to the radiation and, thus, it shrinks in order to conserve the energy. Consequently, the horizon is contracted with respect to its original size and the amount of contraction depends on the energy of the outgoing particle. In this way, the outgoing  particle itself provides the barrier. Now, for the case of a massive BTZ black hole the form of the metric is given by \myref{metric1}, which reduces to a form with \cite{main}
\begin{equation}
f(r)=r^{2}-m-2q^{2}ln(r)-\tilde{M}^{2}c c_{1}r \simeq r^{2}-m-2q^{2}(r-1)-\tilde{M}^{2}c c_{1}r,
\end{equation}
for a constant value of $l=1$. The horizon for this metric is at
\begin{equation}
r_{\pm}=\frac{2q^{2}+\tilde{M}^{2}c c_{1}\pm \sqrt{(2q^{2}+\tilde{M}^{2}c c_{1})^{2}-4(2q^{2}-m)}}{2},
\end{equation}
so that we can write the metric \myref{metric1} in the following modified form
\begin{equation}\label{ds2}
ds^2=-(r^{2}-m-2q^{2}(r-1)-\tilde{M}^{2}c c_{1}r)dt^2+(r^{2}-m-2q^{2}(r-1)-\tilde{M}^{2}c c_{1}r)^{-1}dr^2+r^2d\phi^2.
\end{equation}
Notice that, at $r_+$ there is a coordinate singularity, therefore, in order to study the physics across the horizon we need to change the coordinate system again such that the metric is well behaved at the horizon. Therefore, we use the  Painlev\'e time $t$ \cite{painleve}, which defines a new time coordinate with respect to the Schwarzschild time $t_s$ with an arbitrary function $\tilde{R}(r)$
\begin{align}
t&=t_s-\tilde{R}(r),\quad dt=dt_s-\tilde{R}'(r)dr,\notag\\
dt^2&=dt^{2}_{s}+\tilde{R}'^2(r)dr^2-2\tilde{R}'(r)drdt_s,
\end{align}
with which we can rewrite \myref{ds2} as follows
\begin{equation}\label{ds21}
ds^2=-f(r)dt^{2}_{s}+\left[f^{-1}(r)-f(r)\tilde{R}'^2(r)\right]dr^2+2f(r)\tilde{R}'(r)drdt_{s}+r^2d\phi^2.
\end{equation}
Since, $\tilde{R}(r)$ has been considered as an arbitrary function, we have the freedom to specify it in such a way that the coefficient of $dr^2$ in \myref{ds21} becomes unity and, thus, $\tilde{R}'(r)=\sqrt{1-f(r)}/f(r)$, so that \myref{ds21} further reduces to
\begin{equation}
ds^2=-f(r)dt^{2}_{s}+2\sqrt{1-f(r)}drdt_{s}+dr^2+r^2d\phi^2.
\end{equation}
Correspondingly, the radial null geodesic is given by 
\begin{align}
0&=-f(r)dt^{2}_{s}+2\sqrt{1-f(r)}drdt_{s}+dr^2\notag \\
0&=-f(r)+2\sqrt{1-f(r)}\frac{dr}{dt_{s}}+\left(\frac{dr}{dt_{s}}\right)^2,
\end{align}
so that 
\begin{equation}\label{NullGeo}
\dot{r}=\pm 1-\sqrt{1-f(r)},
\end{equation}
where the upper (lower) sign corresponds to the outgoing (ingoing) geodesics with the assumption that the time increases towards the future. Let us now consider the pair production inside the horizon at $r_{in}\simeq r_{+}$. If, $\omega$ be the energy of the  particle created, the mass of the black hole after the emission of the particle becomes $m-\omega$ and, hence, the horizon contracts from $r_{in}=\big[2q^{2}+\tilde{M}^{2}cc_1 + \sqrt{(2q^{2}+\tilde{M}^{2}cc_1)^{2}-4(2q^{2}-m)}\big]/2$ to $r_{out}= \big[2q^{2}+\tilde{M}^{2}cc_1 + \sqrt{(2q^{2}+\tilde{M}^{2}cc_1)^{2}-4(2q^{2}-m+\omega)}\big]/2$. The difference between $r_{out}$ and $r_{in}$ acts as a barrier of potential $\mathcal{V}$ for the particle tunneling. In this region, $\omega < \mathcal{V}$ and, therefore, the action is imaginary which can be written as follows 
\begin{equation}
\text{Im}\mathfrak{S} = \text{Im}\int_{r_{in}}^{r_{out}} p_{r} dr = \text{Im}\int_{r_{in}}^{r_{out}} \int_{0}^{p_{r}}dp'_{r} dr= \text{Im}\int_{r_{in}}^{r_{out}}\int_{m}^{m-\omega}\frac{dH}{\dot{r}} dr,
\end{equation}
where we use the Hamilton's equation to replace $dp'_{r}$ by $dH/\dot{r}$, followed by a change of variable from momentum to energy. Subsequently, by considering the case of outgoing geodesic in \myref{NullGeo} we obtain
\begin{alignat}{1}
\text{Im}\mathfrak{S} &=\text{Im}\int_{r_{in}}^{r_{out}}\int_{0}^{\omega}\frac{-d\omega'}{1-\sqrt{1-f(r)}} dr \\
&= \text{Im} \int_{r_{in}}^{r_{out}}\int_{0}^{\omega}\frac{-d\omega'}{1-\sqrt{1-r^{2}+m+2q^{2}(r-1)+\tilde{M}^{2}cc_1r-\omega'}} dr, \label{Eq312}
\end{alignat}
where $H=m-\omega'$. Note that while the self-gravitation of the system is taken into account, the mass of the black hole decreases from $m$ to $m-\omega'$ and, thus, we replace $m$ by $m-\omega'$ in \myref{Eq312}. Now, considering $u=1-r^{2}+m+2q^{2}(r-1)+\tilde{M}^{2}cc_1r-\omega'$, we have $du=-d\omega'$, therefore, we can write 
\begin{align}\label{11}
\text{Im}\mathfrak{S} &=Im \int_{r_{in}}^{r_{out}}\int_{u(0)}^{u(\omega)}\frac{du}{(1-\sqrt{u})} dr,
\end{align}
which has a simple pole at $u=1$ and, thus, the residue at $u=1$ is $-2$. Therefore, \myref{11} becomes
\begin{equation}
\text{Im}\mathfrak{S} =-\text{Im} \int_{r_{in}}^{r_{out}}4\pi dr =-4\pi(r_{out}-r_{in}).
\end{equation}
Correspondingly, the transmission probability is given by
\begin{alignat}{1}\label{4}
\Gamma(\omega)&\simeq e^{-2 \text{Im}\mathfrak{S}}= e^{8\pi(r_{out}-r_{in})}= e^{8\pi\sigma},
\end{alignat}
with
\begin{align}
\sigma=r_{out}-r_{in}&=\frac{-4\omega}{\sqrt{(2q^{2}+\tilde{M}^{2}cc_1)^{2}-4(2q^{2}-m)}}+\frac{8\omega^{2}}{\left[(2q^{2}+\tilde{M}^{2}cc_1)^{2}-4(2q^{2}-m)\right]^{3/2}},
\end{align}
where we have considered the binomial series upto second order in $\omega$. The second term in the exponential in \myref{4} is the non-thermal correction to Hawking radiation, whereas the first order term corresponds to the Boltzmann factor $exp[-\frac{\omega}{T}]$, such that the Hawking temperature $T_H$ turns out to be
\begin{align}
T_{H}=\frac{\sqrt{(2q^{2}+\tilde{M}^{2}cc_1)^{2}-4(2q^{2}-m)}}{32\pi }=\frac{2r_{+}-2q^2-\tilde{M}^{2}cc_1}{32\pi},
\end{align}
where the effect of the massive parameter is clearly visible.
\section{Black Hole Chemistry}\label{sec4}
It is customary that every black hole parameter is associated with a chemical equivalent compatible with the first law of thermodynamics \cite{Frassino:2015oca,Kubiznak:2016qmn}. Therefore, we can write
\begin{align}
dE&=TdS+VdP+ \text{work terms},\\
dM&=\frac{\kappa}{8\pi}dA+\Omega dJ+\Phi dq,
\end{align}
and compare the mass $M$ with the internal energy $E$, surface gravity $\kappa$ with the temperature $T$ and the horizon area $A$ with entropy $S$, however, we do not have any gravitational analogue for pressure $P$ and volume $V$ in space-time with $\Lambda=0$. But, for space-time with non-zero cosmological constant it is possible to find an analogue to pressure-volume terms. The basic idea of black hole chemistry is to regard $\Lambda$ as a thermodynamical variable in analogy to the pressure in the first law. The mass $M$ is then considered to be the gravitational analogue of chemical enthalpy, which we denote by $M'$. Under this framework, the pressure $P$ is related to cosmological constant $\Lambda$ as
\begin{align}\label{c}
P&=-\frac{\Lambda}{8\pi}=\frac{(D-2)(D-1)}{16\pi l^{2}},
\end{align}
where $D$ is the dimension of the system. The most general Smarr formula \cite{LSmarr} for $D< 4$ for a charged singly-rotating black hole is given by \cite{Frassino:2015oca,Kubiznak:2016qmn}
\begin{align}\label{a}
(D-3)G_{D}M&=(D-2)TS+(D-2)\Omega J-2VP+(D-3)\Phi q,
\end{align}
where $J$ is the angular momentum, $\Omega$ represents angular velocity and $G_{D}$ stands for the $D$-dimensional Newton's constant. For the case of charged non-massive black holes, \myref{a} and the first law of thermodynamics 
\begin{align}\label{b}
dM'=TdS+VdP+\Phi dq,
\end{align}
hold \cite{Frassino:2015oca}, however, our motivation is to test whether both of them are satisfied for the charged massive BTZ black hole with mass term $\tilde{M}$. Since, in this case, the metric is given by \myref{metric1}, we obtain the temperature as follows \cite{Frassino:2015oca}
\begin{align}
T&=\frac{f'(r_{+})}{2\pi}=\frac{r_{+}}{\pi l^{2}}-\frac{q^{2}}{\pi r_{+}}-\frac{\tilde{M}^{2}cc_1}{2\pi}.
\end{align}
We also compute the entropy $S$, pressure $P$ and enthalpy $M'$ as given in the following
\begin{alignat}{1}
S&=\frac{1}{2}\pi r_{+}, \quad P=\frac{1}{8\pi l^{2}}, \label{Eq47}\\
 M'&=\frac{r^{2}_{+}}{4l^{2}}-\frac{q^2}{2}\ln \bigg(\frac{r_{+}}{l}\bigg)-\frac{\tilde{M}^{2}cc_1r_{+}}{4}. \label{Eq471}
\end{alignat}
By using the above equations \myref{Eq47} and \myref{Eq471}, we can rewrite the enthalpy as
\begin{equation}\label{dq}
M'(S,P,q)=\frac{8S^{2}P}{\pi}-\frac{q^{2}}{4}\ln \bigg(\frac{32PS^{2}}{\pi}\bigg)-4l^{2}\tilde{M}^{2}cc_1SP,
\end{equation}
so that the volume $V$ and the electric potential $\Phi$ turn out to be 
\begin{eqnarray}
V &=& \frac{dM'}{dP}\bigg|_{(S,q)}=2\pi r_{+}^{2}-2q^{2}\pi l^{2} -2\pi r_{+}l^{2}\tilde{M}^{2}cc_1, \label{theV}\\
\Phi &=& \frac{dM'}{dq}\bigg|_{(S,P)}=-q\ln \bigg(\frac{r_{+}}{l}\bigg). \label{ElecP}
\end{eqnarray}
Note that, the volume defined by \myref{theV} is a thermodynamical volume and is not the usual geometrical volume. Nevertheless, it is easy to cross check that all of our results satisfy the first law of thermodynamics \myref{b}, which ensure the fact that all of our calculations are indeed correct. However, the Smarr relation \myref{a} is not satisfied in the given case as expected and it is indicated already in some articles, for instance in \cite{Frassino:2015oca,main}, that for massive black holes the usual Smarr relation may be violated and one may need to modify it accordingly. In order to preserve the Smarr relation let us introduce an extra parameter $\mathcal{G}$ in the Enthalpy $M'$ \myref{Eq471} as follows
\begin{equation} \label{EnthalNew}
M'=\frac{r^{2}_{+}}{4l^{2}}-\frac{q^2}{2}\ln \bigg(\frac{r_{+}}{l}\bigg)-\frac{\tilde{M}^{2}cc_1r_{+}}{4\mathcal{G}}=\frac{8S^{2}P}{\pi}-\frac{q^{2}}{4}\ln \bigg(\frac{32PS^{2}}{\pi}\bigg)-\frac{4l^{2}\tilde{M}^{2}cc_1SP}{\mathcal{G}},
\end{equation}
so that the electric potential remains the same as given by \myref{ElecP}, however, the volume is modified as
\begin{equation}\label{VolumeNew}
V=\frac{dM'}{dP}\bigg|_{(S,q,\mathcal{G})}=2\pi r_{+}^{2}-2q^{2}\pi l^{2} -\frac{2\pi r_{+}l^{2}\tilde{M}^{2}cc_1}{\mathcal{G}}=\frac{8S^2}{\pi}-\frac{q^2}{4P}-\frac{\tilde{M}^2cc_1S}{2\pi P\mathcal{G}}.
\end{equation}
Because of the insertion of $\mathcal{G}$, we are forced to introduce another parameter $K$
\begin{equation}\label{KK}
K=\frac{dM'}{d\mathcal{G}}\bigg|_{(S,P,q)}=\frac{\tilde{M}^2cc_1r_+}{4\mathcal{G}^2},
\end{equation}
which is the thermodynamic conjugate to $\mathcal{G}$. In order to conserve the first law of thermodynamics,  \myref{EnthalNew} enforces us to introduce the parameter $\mathcal{G}$ into the metric \myref{metric1} also, so that \myref{fr} is modified to
\begin{equation}
f(r)=-\Lambda r^2-m-2q^{2}\ln{(\frac{r}{l})}+\frac{\tilde{M}^{2}c c_{1} r}{\mathcal{G}},
\end{equation}
and, thus, the temperature obtains a new form
\begin{equation}\label{TT}
T=\frac{f'(r_{+})}{2\pi}=\frac{r_{+}}{\pi l^{2}}-\frac{q^{2}}{\pi r_{+}}-\frac{\tilde{M}^{2}cc_1}{2\pi\mathcal{G}}.
\end{equation}
With these new formalism, we propose a generalization of the Smarr formula \myref{a}
\begin{equation}\label{SmarrNew}
(D-3)G_{D}M=(D-2)TS+(D-2)\Omega J-2VP-2K\mathcal{G}+(D-3)\Phi q,
\end{equation}
which can be applied to the BTZ black holes including the massive case in $D<4$. It is straightforward to verify that the generalized Smarr relation \myref{SmarrNew} is satisfied with the expressions given in \myref{Eq47}, \myref{EnthalNew}, \myref{VolumeNew}, \myref{KK} and \myref{TT}. Surely, the new results satisfy the first law of thermodynamics \myref{b}, which ensures that there is no violation of the basic principle even after introducing the new parameter $\mathcal{G}$. Let us now set out to calculate the Gibbs free energy and its derivative with respect to $r_+$
\begin{eqnarray}
&& G=M'-TS=-\frac{r^{2}_{+}}{4l^{2}}-\frac{q^2}{2}\ln \bigg(\frac{r_{+}}{l}\bigg)+\frac{q^{2}}{2}, \label{Gibbs}\\
&& \frac{dG}{dr_{+}}=-\frac{r^{2}_{+}+q^{2}l^{2}}{2r_{+}l^{2}} <0. \label{Derivative}
\end{eqnarray}
Note that the Gibbs free energy $G$ does not have any additional effect arising from the new parameter $\mathcal{G}$, viz., \myref{Gibbs} remains invariant with any of the formalisms considered, and so the second equation \myref{Derivative}. In rest of the article we shall only explore the case where the parameter $\mathcal{G}$ is considered, since the other case does not satisfy the Smarr relation and we consider that case to be non-physical. Nevertheless, the second equation \myref{Derivative} implies that the massive charged BTZ black hole does not admit any critical Van der Waals behavior. While verifying the validity of the Reverse Isoperimetric Inequality \cite{Frassino:2015oca}, we find
\begin{equation}\label{Inequality}
\mathfrak{R}=\frac{2\pi}{A}\sqrt{\frac{V}{2\pi}}=\sqrt{1-\frac{q^{2}l^{2}}{r_{+}^{2}}-\frac{l^{2}\tilde{M}^{2}cc_1}{\mathcal{G}r_{+}}} < 1, \qquad\text{with}~ A=4S,
\end{equation}
which means that in our case for all non-zero $q$ and $\tilde{M}$ the Reverse Isoperimetric Inequality \myref{Inequality} is violated  and, hence, the entropy exceeds the expected thermodynamic maximum giving rise to a super-entropic black hole. People sometimes try to restore the Reverse Isoperimetric Inequality by introducing a new thermodynamical variable, for instance, in \cite{Frassino:2015oca} the authors include a renormalization length scale to make the isoperimetric ratio $\mathfrak{R}\geq 1$. However, in our case we notice that because of the presence of the massive parameter $\tilde{M}$ it is never possible to make $\mathfrak{R}\geq 1$ and, thus, the BTZ black hole in massive gravity remains super-entropic always. It should be noted that the super-entropic solution is sometimes even more interesting than the ordinary AdS black holes satisfying Reverse Isoperimetric Inequality. They give rise to non-compact event horizons with finite area, for further details on the interesting facts about the super-entropic black holes, one may see, for instance \cite{Hennigar:2014cfa}.
\section{ Heat Engines}\label{sec5}
In this section, we explore the behavior of a massive charged BTZ black hole \myref{a} as a heat engine \cite{Hennigar:2017apu,Johnson:2014yja}. Our main interest is to calculate the efficiency $\eta$ of the cycle
\begin{align}
\eta&=\frac{W}{Q_{H}}=1-\frac{Q_{C}}{Q_{H}},
\end{align}
where $W$ is the net output work and $Q_{H},Q_{C}$ are the net input and output heat flow, respectively. We will now consider a rectangular cycle in the $P-V$ plane, which is familiar as the Carnot cycle. This rectangle will be in $(P,V)$ coordinates with the corner points described by $(P_{i},V_{j})$, where $i=T,B$ stand for top and bottom and $j=L,R$ denote the left and right. The efficiency of the Carnot cycle is \cite{Hennigar:2017apu} 
\begin{align}\label{efficiency}
\eta&=\frac{W}{Q_{H}}=\frac{(P_{T}-P_{B})(V_{R}-V_{L})}{M'(P_{T},V_{R})-M'(P_{B},V_{L})-V_{L}(P_{T}-P_{B})}.
\end{align}
The total amount of work done is given by the area enclosed by the rectangle. As we see that $S$ in \myref{Eq47} and $V$ in \myref{VolumeNew} are not independent, thus, the adiabats and isochores are the same \cite{Johnson:2014yja}. Consequently, the Carnot and the Stirling cycles coincide. Nevertheless, in order to compute the efficiency \myref{efficiency}, we first solve $S$ from \myref{VolumeNew} as $S=(\tilde{M}^2cc_1\pm A_1)/(32P\mathcal{G})$, with $A_1(P,V)=\sqrt{\tilde{M}^4c^2c_1^2+32\pi P\mathcal{G}(q^2+4PV)}$ and then substitute the solution in \myref{EnthalNew}, so that we obtain the enthalpy as
\begin{equation}\label{MP}
M'(P,V)=PV+\frac{q^2}{4}\left\{1+ln(32\pi P)+2ln\left(\frac{\mathcal{G}}{A_1\pm\tilde{M}^2cc_1}\right)\right\}\mp\frac{\tilde{M}^2cc_1\left(A_1\pm\tilde{M}^2cc_1\right)\left(8\pi Pl^2\mathcal{G}-1\right)}{64\pi P\mathcal{G}^2}.
\end{equation}
Now, $M'(P_{T},V_{R})$ and $M'(P_{B},V_{L})$ can be obtained by replacing $P$ by $P_T,P_B$ and $V$ by $V_R,V_L$ in \myref{MP}, respectively, so that we can compute the efficiency \myref{efficiency} as
\begin{equation}\label{qw}
\eta=\frac{64\pi P_TP_B\mathcal{G}^2(P_{T}-P_{B})(V_{L}-V_{R})}{\mathcal{N}},
\end{equation}
with
\begin{alignat}{1} \label{qw1}
\mathcal{N}=16\pi P_TP_B\mathcal{G}^2 & \left[4P_T(V_L-V_R)+q^2ln\left\{\frac{P_B(A_2\pm\tilde{M}^2cc_1)^2}{P_T(A_3\pm\tilde{M}^2cc_1)^2}\right\}\right]+\tilde{M}^4c^2c_1^2(P_T-P_B) \notag \\
& \qquad\qquad\qquad\quad +\tilde{M}^2cc_1\left\{P_B\left[\pm 8\pi l^2P_T\mathcal{G}(A_2-A_3)\mp A_2\right]\pm A_3P_T\right\},
\end{alignat}
where $A_2=A_1(P_T,V_R)$ and $A_3=A_1(P_B,V_L)$. Therefore, we notice that the mass term $\tilde{M}$ has a significant effect on the efficiency of the heat engine. The form of the efficiency in \myref{qw} along with \myref{qw1} stands for a complete general expression for the massive BTZ black hole, which one can analyze further. 
\section{Concluding remarks}
We have studied a black hole solution of a charged massive BTZ black hole by using the Vainshtein and dRGT mechanism and analyzed the effects of such black hole in different scenarios. In particular, we have studied the Hawking radiation form this solution by utilizing the tunneling formalism. The solution that we obtain violates the Reverse Isoperimetric Inequality and, thus, they are super-entropic. They also do not admit any critical Van der Walls behavior. Moreover, we have explored the black hole chemistry followed by the efficiency of a heat engine from the $P-V$ diagram for this system. Our results show that unlike the ordinary BTZ black holes, the thermal fluctuation is absent in the massive gravity case. Thus, the BTZ black holes in massive gravity scenario are free from instabilities which is perhaps one of the greatest shortcomings for ordinary BTZ black holes. Furthermore, since the thermal fluctuations may be interpreted as quantum effects, the massive gravity does not feel important quantum effects.

There are many interesting aspects which can be followed from our results. Firstly, it would be interesting to analyze the holographic entanglement entropy and holographic complexity dual to the BTZ AdS black hole. Secondly, the holographic conductivity has also been analyzed in certain aspects of massive gravity \cite{conduct}. Therefore, it would be interesting to analyze such effects in our case also. Finally, our solution can be used to construct a CFT dual by using the standard formulation of AdS/CFT correspondence \cite{ads, M3, M4, M5, cft}. \\

\noindent \textbf{\large{Acknowledgements:}} S.D. acknowledges the support of research grant (DST/INSPIRE/04/2016/ 001391) from the Department of Science and Technology, Govt. of India.



\end{document}